\documentclass[12pt]{article}
\begin{document}
\title{Quantum Behavior in Machian Model}
\author{{\bf Merab Gogberashvili}\\
Andronikashvili Institute of Physics \\
6 Tamarashvili Street, Tbilisi 0177, Georgia \\
and \\
Javakhishvili State University\\
3 Chavchavadze Avenue, Tbilisi 0128, Georgia  \\
\\
{\sl E-mail: gogber@gmail.com }} 
\maketitle
\begin{abstract}
The essentials of quantum theory, the Schr\"odinger equation and the Planck constant, are derived using classical statistical mechanics within the non-local Machan model. The appearance of complex wave function is connected with the necessity of quantization condition when one neglects the existence of the preferred frame and assumes Galilean symmetry of the model.
\vskip 0.5cm
PACS numbers: 03.65.Ta; 05.20.Gg; 45.20.-d; 04.50.Kd
\end{abstract}

\vskip 0.5cm


Although the predictions of quantum mechanics do not contradict relativity there is a conceptual gap between these theories and many physicist believe in existence of some deterministic causes for quantum phenomena (see, for example \cite{hidden} and references therein). This point of view is also reinforced by experiences with pseudo-random behavior in classical statistical mechanics. The well-known difficulty of hidden-variable models is that the assumption of existence objective elements of reality even without measurements leads to non-locality \cite{NonLoc}. However, an essential ingredient of the deterministic quantum mechanics concerns the fact that our universe evolves. Indeed, the usual interpretation of quantum mechanics, which implies that nature is fundamentally random, raises the questions like: Was the universe existed before human beings appear on the Earth? 

One example of non-local approach is Machian model of gravity \cite{Gog1,Gog2}, which assumes that in any two particle gravitational interaction all particles of the universe are involved. These interactions create a fundamental cosmological preferred frame, which may be detected even locally using CMB. In this picture geometry is not an independent entity but derivative from the properties of matter in 'gravitationally entangled' universe, i.e. a concise way to describe the symmetries of the system in IR. Some consequences of the Machian approach are: The relativistic effects pertain only to matter fields, while gravitation in UV should single out a cosmological frame and thus remain non-relativistic; Minkowski space-time corresponds to the case of the isotropic and homogeneous universe and not to the empty vacuum with zero energy as in general relativity; The speed of light and Newton's constant both become effective quantities and depend on the matter content of the universe. However, the Machian model in IR imitate some basic features of the special and general relativity theories, such as weak equivalence principle and local Lorentz invariance for matter field. So the model obeys so-called signal locality, i.e. does not allow matter fields to move faster than light. 
 
To fix parameters of the Machian model it was considered the spherical model-universe of the radius $R$ consisting of the ensemble of $N$ uniformly distributed identical particles of gravitational mass $m_g$ \cite{Gog2,Bar}. The Newtonian 'universal' gravitational potential, which acts on any particle from the ensemble, 
\begin{equation} \label{Phi-N2}
\Phi \approx - \frac{m_g G}{R}~N^2 \approx - \frac{M_g G}{R} ~,
\end{equation}
($G$ is the Newton constant) is constant since for the horizon distances, $R$, the universe is isotropic and homogeneous (the cosmological principle). Due to the non-local interactions the active gravitational mass of the universe is equal to
\begin{equation} \label{M}
M_g \approx N^2 m_g~,
\end{equation}
and not $\sim N m_g$, as expected from the Gauss law in the additive case. The value of $M_g$ (that includes all dark components as well) can be found from the critical density condition, which is valid even in Newtonian cosmology \cite{Bar}. The radius of the universe, $R$, or the radius of the causal sphere of any object, can be estimated using the experimental values of the Hubble constant and the speed of light. So for the parameters entering (\ref{Phi-N2}) we have,
\begin{equation} \label{MRG}
M_G \sim 10^{53}~ kg, ~~~ R \sim 10^{26}~m~, ~~~ G \approx 6.7~ 10^{-11} \frac{m^3}{s^2 kg}~. 
\end{equation}
Then the value of the 'universal' potential $\Phi$, one of the main parameters of the model, numerically equals to the square of light speed \cite{Gog1,Gog2,Sia}, 
\begin{equation} \label{Phi}
\Phi = - \frac{M_g G}{R} = - c^2 ~.
\end{equation}
Using this estimation the Machian energy of a particle,
\begin{equation} \label{E_0}
E = - m_g \Phi ~,
\end{equation}
recovers Einstein's famous energy-mass relation.

We can estimate also the amount of gravitating particles $N$ within the Hubble horizon. The mass of a typical particle, for example a proton, has the order $m_p \sim 10^{-27} kg$. From (\ref{M}) we know that the gravitational mass of the universe $M_g$ is $N$ times larger than the sum of the masses of all particles. So the proton equivalent of the total number of gravitating particles is
\begin{equation} \label{N}
N \sim \sqrt{\frac{M_g}{m_p}} \sim 10^{40}~.
\end{equation}
This number, the another main parameter of the model, appears in different context in the Dirac large numbers hypothesis proving the existence of some deep connection between micro and macro physics \cite{large}. 

The assumption of involving the universe in local interactions effectively weakens the observed scale of gravity since, due to non-local interactions, gravitational mass of a particle (the 'charge' of gravity) appears to be much larger than its inertial mass, 
\begin{equation} \label{m=Nm}
m_g = \sqrt{N} m~.
\end{equation}
This assumption does not mean a violation of the weak equivalence principle, which is well tested \cite{EP}, since all results of standard physics will remain the same if together with (\ref{m=Nm}) we introduce the new gravitational constant,
\begin{equation} \label{g=NG}
g = N G ~.
\end{equation}
where $1/\sqrt{G} \sim 10^{19}~ GeV$ and $1/\sqrt{g} \sim 1 ~TeV$ correspond to so-called Planck and Higgs scales respectively. The relation (\ref{g=NG}) can be addressed to the hierarchy problem in particle physics \cite{Gog2}.

In this paper we want to find equations of motion of particles within the Machian model \cite{Gog1,Gog2}. Because of the existence of non-locality it is natural to consider all particles in the universe as a single statistical ensemble and try to describe those using laws of the statistical mechanics. Due to huge size of the universe this assumption seems unnatural. However, it is known that number of physical systems can exhibit similar behaviors at very different length scales. For example, the laws of hydrodynamics are successfully applied even in cosmology. 

We shall demonstrate below that the classical equations of motion for particles in our non-local model are equivalent to the Schr\"odinger equation. In standard quantum theory the Schr\"odinger equation is postulated. However, there exist numerous attempts to strictly derive it from classical physics. For example, using the hydrodynamical interpretation of quantum mechanics \cite{Mad}, in the Nelson theory of stochastic mechanics \cite{Nel}, within the hidden-variable theories of de Broglie and Bohm \cite{Bohm}, with 'exact uncertainty principle' \cite{Ha-Re}, or by non-equilibrium thermodynamics \cite{Gro}. 

The description of physical systems by ensembles may be introduced at quite a fundamental level using notations of probability and action principle. We start with using of the Machian energy of a particle (\ref{E_0}) to define its action \cite{Gog1},
\begin{equation} \label{A}
A = -\int_{t_1}^{t_2} dt~ E \approx - m_g c^2 (t_2 - t_1) \approx - m_g c r ~,
\end{equation}
where $r \approx R/N $ is the mean distance between particles from the ensemble. If the time moments $t_1$ and $t_2$ correspond to uniform velocities, the universality of Machian energy for inertial frames (our version of the relativity principle) leads to the principle of least action,
\begin{equation} \label{deltaA}
\delta A = 0~.
\end{equation}
Since for the ensemble of particles initial conditions are not known exactly to describe the motion of individual particles we need to introduce some density function $P_n(t)$ ($n = 1,2,...,N$), which takes into account positions of all particles. Then we can write action integral (\ref{A}) in more general form: 
\begin{equation} \label{AP}
A = - \sum_n\int dt~P_n E ~.
\end{equation} 
To interpret $P_n(t)$ as the probability density it must be normalized, 
\begin{equation}
\sum_n P_n (t) = 1 ~. 
\end{equation}
To find the normalization constant let us by (\ref{A}) calculate the maximal action corresponding to the whole universe,
\begin{equation} \label{A_U}
A_U = - M_g T_U c^2 \approx - M_g R c \approx  N^3 A ~,
\end{equation} 
where $T_U \approx R/c$ is the age of the universe. Using the estimations (\ref{MRG}) and (\ref{N}) we find that the portion of the action of the universe for a single member of the ensemble has the order of the Planck constant,
\begin{equation} \label{hbar}
A \sim \frac{A_U}{N^3} \sim - \frac{M_g R c}{N^3} \sim - 10^{-33}~ \frac{m^2 kg}{s} \sim - h ~,
\end{equation}
and $h$ can be used to normalize the function $P_n(t)$. 

The number of particles in the ensemble, (\ref{N}), is huge and it is convenient to use continuous limit replacing $P_n(t)$ with $P(t,x^i)$ ($i = 1,2,3$), where $x^i$ represent space coordinates of particles. Since the function $P(t,x^i)$ should be positive and normalized with $h$ we write it in the form:
\begin{equation} \label{K}
P(t,x^i) = e^{2 K(t,x^i)/\hbar}~,
\end{equation}
where $\hbar = h/2\pi$. Justification of the value of the normalization constant in (\ref{K}) to be exactly equal $2/\hbar$ will be done below. The new density function $K(t,x^i)$ in (\ref{K}) obeys the normalization condition,
\begin{equation}
\int d^3x ~P = \int d^3x ~e^{2 K(t,x^i)/\hbar} = 1 ~. 
\end{equation}

Note that within our non-local model the number of gravitational particles in the universe, (\ref{N}), together with explanation of the hierarchy between the Higgs and the Planck scales, (\ref{g=NG}), gives the correct value for the Planck constant, (\ref{hbar}). The parameter $\hbar$ is additional to (\ref{Phi}) and (\ref{N}) universal parameter of the model. It can be used, for instance, for the alternative to (\ref{E_0}) definition of a particle's energy,
\begin{equation} \label{homega}
E = \hbar \omega ~.
\end{equation}
Thus a particle can be assigned with the 'frequency' $\omega$ that corresponds to some 'dissipative' precesses in the 'gravitationally entangled' universe. While the definition of energy by $\Phi = - c^2$, (\ref{E_0}), is convenient for massive objects, the alternative definition by $\hbar$, (\ref{homega}), is more useful for elementary particles, since in the limit of zero mass (\ref{E_0}) fails. 

Now consider the simplest case when all $N$ particles in the universe are at rest with respect to each other and are described by some density function $P_0(t,x^i)$. For a particle from the ensemble with the Machian energy (\ref{E_0}) one can introduce the Hamilton principle function 
\begin{equation} \label{S0}
S_0 = - Et + const~,
\end{equation}
and write the action integral (\ref{AP}) in the form:
\begin{equation} \label{A0}
A_0 =\int dtd^3x ~P_0 \left(\frac{\partial S_0}{\partial t}+ V\right)~,
\end{equation}
where $V(t,x^i)$ is some local potential acting on the particle. Variation of this action with respect to $S$ and $P$ by fixed end-point variation ($\delta P = \delta S=0$ at the boundaries) leads to the trivial continuity and Hamilton-Jacobi equations,
\begin{equation} \label{S_0}
\frac{\partial P_0}{\partial t} = \frac{\partial S_0}{\partial t} + V = 0 ~. 
\end{equation}

For the case when only one particle moves with respect to the preferred frame the Hamilton principle function (\ref{S0}) is transformed to:
\begin{equation} \label{S}
S = p_ix^i - Et + const ~,
\end{equation}
where the 3-momentum of the particle, $p^i$, is related with $S(t,x^i)$ by the gradient function,
\begin{equation} \label{p}
p^i = \nabla^i S~.
\end{equation}
The action integral (\ref{A0}) in this case has the form:
\begin{equation} \label{Ap}
A_p =\int dtd^3x~ P \left( \frac{\partial S}{\partial t} + \frac{\nabla_i S \nabla^i S}{2m} + V \right)~,
\end{equation}
$m$ is the inertial mass of the particle. Variation of this action gives the system of the Hamilton-Jacobi equation,
\begin{equation} \label{HJE}
\frac{\partial S}{\partial t} + \frac{\nabla_i S \nabla^i S}{2m} + V = 0 ~,
\end{equation}
and the continuity equation, 
\begin{equation} \label{cont}
\frac{\partial P}{\partial t} + \frac{\nabla_i \left( P\nabla^i S\right)}{m} = 0~. 
\end{equation}
Since all other $(N-1)$ particles are at rest we again expect to have the stationary distribution function for the moving particle, i.e. the continuity equation (\ref{cont}) should reduce to:
\begin{equation} \label{u+P}
\frac{\partial P}{\partial t} = \nabla^i S \nabla_i P + P \Delta S = 0 ~,
\end{equation} 
where $\Delta = \nabla^i \nabla_i$. For the case of constant momentum, 
\begin{equation}
\Delta S = \nabla_i p^i = 0~,
\end{equation}
from (\ref{u+P}) we see that there exists some orthogonal to $p^i$ constant 3-momentum, 
\begin{equation}
c^i \sim \frac{\nabla_i P}{P} ~, ~~~~~c^i p_i = 0~.
\end{equation} 
Then the solution of the continuity equation (\ref{cont}) can be written in the form:
\begin{equation} \label{P=P0}
P = P_0 e^{2 \oint_{L} dl_ic^i /\hbar}~,
\end{equation}
where $P_0$ is the distribution function for the case of zero momentum (\ref{S_0}). 

The situation reminds non-equilibrium thermodynamics. The moving particle feels resistance of the ensemble, which can be formalized as if the particle is maintained in a non-equilibrium stead-state by the heat flow $\Delta Q $, i.e.
\begin{equation}
P = P_0 e^{\Delta Q /kT} ~,
\end{equation} 
where $k$ and $T$ are the Boltzmann constant and the absolute temperature respectively. Equating the kinetic energy of the thermostat per degree of freedom, $kT/2$, with the average kinetic energy of an oscillator from the ensemble, $\hbar \omega/2$, and using the Boltzmann relation between heat and action, $\Delta Q = 2\omega S$, together with the description of momentum as a gradient of action function, (\ref{p}), we get (\ref{P=P0}) \cite{Gro}. This analogy 'explains' appearance of $2/\hbar$ in (\ref{K}).

The proportional to $c^i$ factor in (\ref{P=P0}) represents the disturbance of the density function $P(t,x^i)$ of the ensemble due to the motion of one particle. When considering the moving particle to stay with the undisturbed density function, $P_0$, instead of (\ref{S}) one can use the modified Hamilton function
\begin{equation} \label{S'}
S' = \oint_{L} dl_ic^i - E't + const ~,
\end{equation}
that leads to the relation,
\begin{equation} \label{S'P}
\nabla^i S' = \frac \hbar 2 \frac{\nabla^i P}{P}~,
\end{equation}
and automatically 'gauge out' the second term in the continuity equation (\ref{cont}). In terms of the new Hamilton function, (\ref{S'}), the Hamilton-Jacobi equation of our problem, (\ref{HJE}), transforms to:
\begin{equation} \label{HJE'}
\frac{\partial S'}{\partial t} - V_q + V = 0 ~.
\end{equation} 
The second term in this equation is so-called quantum potential,
\begin{equation} \label{Vq}
V_q = -\frac{\hbar^2}{2m} \frac{\Delta P}{P} ~,
\end{equation}
which makes the system of equations (\ref{cont}) and (\ref{HJE'}) inherently non-local. Because of (\ref{S'P}) the change in the density of one particle can affect $V_q$ and hence the motion of correlated particles on any distances. 

It is important that for inertial particles the value of $V_q$ and the form of the Hamilton-Jacobi equations (\ref{HJE'}) are independent of velocity. So in the homogeneous universe there exists a class of privileged observers, having constant velocities with respect to the preferred frame, for which the universe appears spherically symmetric (the standard definition of inertial frames \cite{Wei}). Universality of $V_q$ closely relates to the invariance of Machian energy for inertial observers (this leads to the least action principle (\ref{deltaA}) \cite{Gog1}). To describe this property one can introduce also so-called Fisher information of the probability density \cite{Fish},
\begin{equation} \label{F}
I_F = \int d^3x \frac{\nabla_i P\nabla^i P}{P}~. 
\end{equation}
Connections of (\ref{F}) with the Hamilton principle was noticed in \cite{Hamilton}. 

Since for inertial particles preferred frame is unobservable ($I_F$ is universal) one can try to use the Galilee transformations,
\begin{equation}
t' = t~, ~~~~~x_i' = x_i - v_it~.
\end{equation}
The distributing function in this case, $P_v$, obeys the continuity equation with flow,
\begin{equation} \label{cont-v}
\frac{\partial P_v}{\partial t} + \nabla_i \left( v^i P_v \right) = 0~. 
\end{equation} 
For the inertial particle we again (as in (\ref{u+P})) can restrict ourselves with the stationary state and of 'Hamiltonian flow', which is given by:
\begin{equation} 
\frac{\partial P_v}{\partial t} = \nabla_i v^i = 0~. 
\end{equation}
The action integral for this case takes similar to (\ref{Ap}) form:
\begin{equation} \label{Av}
A_v = \int dtd^3x~ P_v \left( \frac{\partial S_v}{\partial t} + \frac{\nabla_i S_v \nabla^i S_v}{2m} + V \right) ~. 
\end{equation}
The difference of this expression with (\ref{Ap}) is conceptual, now we supposing that all particles in the universe move with some non-zero constant velocities $v^i \sim \nabla^i S_v$. Since the universe in our model is considered to be finite the constant flow will result in change of amount of particles, $N$, and thus the Machian energy, inside the horizon. This requires modification of variational principle by introduction of boundary terms on a closed surface $L$. The boundary terms can be taken into account writing 
\begin{equation} \label{oint}
\oint_L dl^i \nabla_i S_v = n h ~, 
\end{equation}
where $n$ in the number of particles that cross $L$ in unite time. Since $n$ is a integer number this property can be formalized by the introduction of the single valued complex function
\begin{equation} \label{psi}
\psi = e^{i S_v/\hbar} = e^{i \left(S_v/\hbar + 2\pi n\right)}~, 
\end{equation} 
which automatically leads to the relation (\ref{oint}). So the consideration of $S_v$ as a complex phase in (\ref{psi}) express the physical requirement that constant velocities in the Machian model are not equivalent. 

Using the complex Madelung transformation \cite{Mad}
\begin{equation} \label{Psi}
\Psi = \sqrt{P} \psi = e^{(K + i S)/\hbar}~,
\end{equation}
together with the Bohr-Sommerfeld like quantization condition (\ref{oint}) the non-linear system of equations (\ref{HJE}), (\ref{cont}) and (\ref{HJE'}) can be condensed into a single linear Schr\"odinger equation
\begin{equation} \
i \hbar \frac{\partial \Psi}{\partial t} = \left(-\frac {\hbar^2}{2m}\Delta + V \right)\Psi~.
\end{equation}
So within the Machian model the Schr\"odinger equation is a mathematical consequence of the classical statistical mechanics. The advantages of complex formulation are the compactness and the linearity of the Schr\"odinger equation. The Machian approach leads to the interpretation of wave function as an objective-realistic entity and removes disadvantage of traditional formulation of quantum mechanics where the complex wave function $\Psi$ has no direct physical meaning. 

To conclude in this paper the quantum-like properties within the Machian model of gravity \cite{Gog1,Gog2} was considered. The Planck constant was interpreted as a portion of the total action of the finite universe per one gravitating particle. Using classical statistical mechanics, from the non-linear system of continuity and Hamilton-Jacob-like equations, it was derived the Schr\"odinger equation. The crucial moment of the derivation was the introduction of the quantization condition (\ref{oint}) (it is known that classical and quantum Schr\"odinger equations are not equivalent unless (\ref{oint}) is imposed \cite{Wa-Ta}). It was shown that (\ref{oint}) is connected with the existence of the preferred frame in the universe. The fact that Galilean invariance is fictitious symmetry of the classical Schr\"odinger equation, and underline theory should introduce a preferred frame, was already noted in \cite{Val}. The convenience of complex formalism for description of particles in Machian model was directly derived from the assumption that a pair of real classical functions, the probability density and the Hamilton function, which obey the 'single-valuenes' condition, should be associated with any measurement. 


\section*{Acknowledgement:} The author would like to acknowledge the hospitality extended during his visits at the Physics Department of Helsinki University where this work was done.


\end{document}